\documentclass{vldb}
\usepackage{balance}  

\newcommand{\subparagraph}{}

\usepackage{xspace}
\usepackage{verbatim}
\usepackage{tikz}
\usepackage{subcaption}
\usepackage{caption}
\usepackage{graphicx}
\usepackage{titlesec}
\usepackage{amsmath, amsthm, amssymb}
\usepackage{listings}
\usepackage{courier}
\usepackage[T1]{fontenc}

\usepackage{amsmath}
\usepackage{amssymb}
\usepackage{latexsym}
\usepackage{times}
\usepackage{bm}
\usepackage{color}
\usepackage{subscript}
\usepackage{multirow}
\usepackage{booktabs} 
\usepackage{mathtools}
\usepackage{commath}
\usepackage[noend]{algpseudocode}
\usetikzlibrary{arrows}
\usepackage{array}
\usepackage{enumitem}
\usepackage{wrapfig}
\usepackage[utf8]{inputenc} 
\usepackage{url}            
\usepackage{booktabs}       
\usepackage{amsfonts}       
\usepackage{nicefrac}       
\usepackage{microtype}      
\usepackage{cleveref}

\newcommand{\whilec}{\textbf{while}}
\newcommand{\skipc}{\textbf{skip}}
\newcommand{\ifc}{\textbf{if}}

\newcommand{\thenc}{\textbf{then}}
\newcommand{\elsec}{\textbf{else}}
\newcommand{\doc}{\textbf{do}}
\newcommand{\tryc}{\textbf{try}}
\newcommand{\catchc}{\textbf{catch}}

\newcommand{\callc}{\textbf{call}}
\newcommand{\id}[1]{\mathsf{#1}}

\newcommand{\Sy}{\mathsf{Y}}

\newcommand{\sY}{\mathsf{Y}}
\newcommand{\sX}{\mathsf{X}}
\newcommand{\sI}{\mathsf{I}}
\newcommand{\sProg}{\mathsf{Prog}}
\newcommand{\sZ}{\mathsf{Z}}
\newcommand{\M}{\theta}

\newcommand{\eat}[1]{} 

\newcommand{\xp}[2]{P \if*#1\else^{#1}\fi \if*#2\else_{\! #2}\fi}

\long\def \ignoreme#1{}

\def\qed{\hfill \mbox{\rule[0pt]{1.5ex}{1.5ex}}}

\lstnewenvironment{codesmall}
  {\lstset{
        aboveskip=5pt,
        belowskip=5pt,
        escapechar=!,
        mathescape=true,
        language=Java,
        basicstyle=\linespread{0.94}\ttfamily\footnotesize,
        morekeywords={createSet, sendData, makeLambdaFromMember,
                APPLY, FILTER, execute, setInput, setOutput, makeLambda, makeLambdaFromMethod,
                Handle, Vector, Map, pair, makeObjectAllocatorBlock, push_back, Object, makeObject},
        showstringspaces=true}
        \vspace{0pt}%
        \noindent\minipage{0.47\textwidth}}
  {\endminipage\vspace{0pt}}

\lstnewenvironment{codetiny}
  {\lstset{
        aboveskip=5pt,
        belowskip=5pt,
        escapechar=!,
        mathescape=true,
        language=Java,
        basicstyle=\linespread{0.94}\ttfamily\scriptsize,
        morekeywords={createSet, sendData, makeLambdaFromMember,
                APPLY, FILTER, execute, setInput, setOutput, makeLambda, makeLambdaFromMethod,
                Handle, Vector, Map, pair, makeObjectAllocatorBlock, push_back, Object, makeObject},
        showstringspaces=true}
        \vspace{0pt}%
        \noindent\minipage{0.47\textwidth}}
  {\endminipage\vspace{0pt}}
  
\usepackage{stackengine}

\vldbTitle{Searching a Database of Source Codes Using Contextualized Code Search}
\vldbAuthors{Rohan Mukherjee, Swarat Chaudhuri, and Chris Jermaine}
\vldbDOI{https://doi.org/10.14778/3401960.3401972}
\vldbVolume{13}
\vldbNumber{10}
\vldbYear{2020}

\begin{document}
\pagenumbering{gobble}
\captionsetup[figure]{labelfont={bf},labelformat={default},labelsep=colon}
\captionsetup[subfigure]{labelfont={},labelformat={parens},labelsep=space}
\captionsetup[table]{labelfont={bf},labelformat={default},labelsep=colon}


\title{Searching a Database of Source Codes Using Contextualized Code Search}

\numberofauthors{3} 
\author{
%
\alignauthor
Rohan Mukherjee \\
       \affaddr{Rice University}\\
       \affaddr{Houston, USA}\\
       \email{rm38@rice.edu}
\alignauthor
Swarat Chaudhuri \\
       \affaddr{Rice University}\\
       \affaddr{Houston, USA}\\
       \email{swarat@rice.edu}
\alignauthor Chris Jermaine \\
       \affaddr{Rice University}\\
       \affaddr{Houston, USA}\\
       \email{cmj4@rice.edu}
}

\maketitle
\begin{abstract}
Consider the case where a programmer has written some part of a program, but has left part of the program (such as a method or a function body) incomplete.  The goal is to use the context surrounding the missing code to automatically ``figure out'' which of the codes in the database would be useful to the programmer in order to help complete the missing code. The search is ``contextualized'' in the sense that the search engine should use clues in the partially-completed code to figure out which database code is most useful.  The user should not be required to formulate an explicit query.

We cast contextualized code search as a learning problem, where the goal is to learn a distribution function computing the likelihood that each database code completes the program, and propose a neural model for predicting which database code is likely to be most useful.  Because it will be prohibitively expensive to apply a neural model to each code in a database of millions or billions of codes at search time, one of our key technical concerns is ensuring a speedy search. We address this by learning a ``reverse encoder'' that can be used to reduce the problem of evaluating each database code to computing a convolution of two normal distributions.
\end{abstract}

\section{Introduction}

An end-user has produced a partially completed 
computer program,
where a piece of code (typically a body of a method or function) is wholly or partially missing.  The goal is to search a large corpus of 
program fragments $D = \{ \sProg_1, \sProg_2, ... \}$, and automatically choose 
the fragment from the database that is most likely to complete the program, without requiring an end-user to explicitly formulate a query.
For example, consider the following code:

\begin{codesmall}
class IO {
  public void readFully(InputStream fd, 
    byte[] dst, int off, int len) 
                    throws IOException {
      while (len > 0) {
         int r = fd.read(dst, off, len);
         off += r;
         len -= r; }}
 
  public void findMe (OutputStream out){
      __CODE_SEARCH__ }}
\end{codesmall}

\noindent The goal is to find a code in a large database of codes that could replace the missing code indicator.  The 
search is \emph{contextualized} because the user need not describe the search using an explicit query;
the context around the missing code---class member variables, comments, surrounding method signatures,
and so on---is used to power the search. The goal is to provide search functionality ``for free''; the user of an integrated development environment (IDE) need only click on a particular line where the missing code is to be inserted, and the system looks at the partially-completed code and figures out from the context what the answer should be. The user need not take the time to posit
a query, composing keywords or writing an English description of the code that s/he wants.  In this case, one of the top codes returned is:

\begin{codetiny}
/** Writes the contents of this byte array output 
    stream to the specified output stream argument.*/
public void writeTo(OutputStream out) 
                 throws IOException {
  ByteString[] cachedFlushBuffers;
  byte[] cachedBuffer; int cachedBufferPos;
  synchronized (this) {
    cachedFlushBuffers=flushedBuffers.toArray(
        new ByteString[flushedBuffers.size()]);
    cachedBuffer=buffer;
    cachedBufferPos=bufferPos; }
  for(ByteString byteString:cachedFlushBuffers){
    byteString.writeTo(out);
    }
  out.write(copyArray(
            cachedBuffer,cachedBufferPos));
}
\end{codetiny}

\noindent The system was able to infer from the surrounding class---which included a method that reads from an input stream---that the user was looking for a \texttt{write} method.  


\vspace{8 pt}
\noindent
\textbf{Code search via neural embedding.} 
Methods for learning neural embeddings have become widespread.  The idea is to learn a neural function that is able to map objects to a position in a high-dimensional space, such that objects that are similar or related are positioned closely to one another.  Methods for computing word embeddings such as Word2Vec \cite{word2vec} and BERT \cite{bert} are the best examples of this.  Not surprisingly, such methods have been applied to code search, especially for powering natural language-based search \cite{facebook_mapl18, deep_code_search,attention-deep-search-paper}. The idea is to learn one neural function that embeds a query in a high-dimensional space, and another that embeds a code in the same space.  Code search then reduces to nearest-neighbor search.

Unfortunately, there may be little reason to believe that such methods will work for contextualized code search (CCS). In CCS, both queries and codes are exceedingly complex objects, so that learning a generalizable embedding from millions of query-code co-occurrences in a training set seems hard. Queries in contextualized code search are inherently multi-modal, with a large number of disparate evidences as 
to what the query result should
be: sets of types, method
calls and keywords surrounding the missing code, natural language comments, sequences of formal parameters, and so on.  Because of this, each query in a training corpus is likely unique, and will only be seen once in the training data. Compare this to the problem of learning a word embedding, where each word is seen many times in many different contexts.  As a result, over-fitting may be a significant problem, leading to poor search performance.

\vspace{8 pt}
\noindent
\textbf{Code search as program synthesis.}
We propose a unique approach to CCS, where we view CCS as a special case of statistical program synthesis.  Statistical program synthesis \cite{manna_synthesis, DeepCoder, bayou} is the problem of learning how to automatically write programs.  In particular, we view CCS as a variety of \emph{conditional program generation} \cite{bayou}, where a learner learns to use the context $\sX$ collected from the surrounding code to realize a posterior distribution function $P(\sProg|\sX) = \int_{\sZ} P(\sProg | \sZ) P(\sZ | \sX) d\sZ$, for a latent variable $Z$\footnote{In the paper, we will use the convention that a mathematical object written in a sans-serif font such as $\sX$ represents an observed value, while an italicized object such as $X$ represents a random variable. Hence, $P(X)$ refers to the distribution of random variable $X$, while $P(\sX)$ refers to the likelihood of observing value $\sX$ for random variable $X$.}.  $Z$ can be viewed as an unknown specification for the code to be generated.  When treating CCS as an instance of conditional program generation, search is the task of finding the database program such that $\sProg$ $=$ $\textrm{argmax}_{\sProg' \in D} P(\sProg' | \sX)$.

In contrast to more traditional, embedding-based approaches that attempt to map both context and program to similar representations in a high-dimensional latent space, conditional program generation attempts to learn to generate the program from the context.  This may be more resistant to over-fitting because the statistical program synthesizer must learn to accurately generate $\sProg$, despite of the uncertainty in $\sZ$ embodied by the distribution $P(Z | \sX)$.

The key difficulty with re-casting contextualized code search as synthesis is computational.
When the goal is to synthesize a program by generating $\sProg$ so as to maximize
$P(\sProg | \sX)$, seconds or even minutes of compute time can be devoted to solving the resulting optimization problem.  
During search, however, for a given query evidence $\sX$, $P(\sProg_i | \sX)$ must be evaluated for millions or
billions of values of $i$ very quickly, while the user waits.  In a typical implementation, $\sProg_i$ will 
take the form of a parse tree, and 
evaluating $P(\sProg_i | \sX)$ requires repeatedly
pushing productions in that parse three through a neural network.  Doing this quickly for millions of different programs
will not be feasible.  
To address this, 
we force the posterior $P(Z | \sX)$ distribution over the unknown specification $Z$ to be multivariate Gaussian.  When learning 
$P(\sProg | \sX)$, we concurrently learn an approximation $Q(Z | \sProg) \approx P(Z | \sProg)$ 
(a so-called ``reverse encoder'') 
where
$Q(Z | \sProg)$ is also constrained to be normal.  Computing $P(\sProg | \sX)$ then reduces to computing
a convolution of  $P(Z | \sX)$ and $Q(Z | \sProg)$, which is computationally trivial when both distributions are multivariate Gaussian,
leading to a very fast search.

\vspace{5 pt}
\noindent
\textbf{Our contributions.}
Key contributions of our work are:
\begin{itemize}[leftmargin=*]
\vspace{-5 pt}

\item We introduce the problem of contextualized code search. While code search has been studied for a long time (see Section 3 of the paper), prior efforts have typically been powered by user-supplied queries. In contrast, in CCS, the query is implicit, and inferred by the surrounding context. We are the first to study code search using this type of implicit query.

\vspace{-5 pt}
\item We present a unified probabilistic framework in which a disparate, multi-modal set of contextual evidences $\sX$ can be synthesized into
a posterior distribution $P(\sZ | \sX)$ over the unknown specification for the code being searched for.  This distribution
encodes the uncertainty inherent in search.

\vspace{-5 pt}
\item We consider how to design the learning problem to ensure that search can happen quickly. 

\vspace{-5 pt}
\item Finally, we experimentally evaluate our tool for CCS (called  \textsc{Codec}) over a corpus consisting of nearly one billion lines of code. We show experimentally that a 16-GPU machine can be used to search our database size of 27.9 million Java methods in little over a second.
\end{itemize}


\section{Example Application}

In this section, we give a more detailed example of CCS, via a short case study that demonstrates our tool, called \textsc{Codec} (Contextualized cODe sEarCh).
We call our system \textsc{Codec} to emphasize the synthesis-based approach to code search: the system learns to encode the context and decode that encoding into a program, rather than simply learning to encode contexts and programs into a latent space.
\vspace{5 pt}

Consider the following unfinished user interface code:

\begin{codesmall}
import javax.swing.*;
class MyGuiAppl{

  /**
  create a new frame
  */
  public JFrame ?(? a){
    __CODE_SEARCH__; }}
\end{codesmall}
\vspace{5 pt}

\noindent
\textsc{Codec} extracts the class name \texttt{MyGuiAppl}, the Javadoc text for the method with the missing body (``\texttt{create a new frame}''), as well as the desired return type (\texttt{JFrame}) and the name of the formal parameter (\texttt{a}). Since no method name and no formal parameter type are given, these are ignored. \textsc{Codec} searches a database of code fragments and returns the following code in its top few results:

\vspace{5 pt}
\begin{codesmall}
/** 
 * Creates a new UserInterface object.
 * @param title the title
 * @return the j frame
 */
public JFrame createFrame(final String 
  title){
  JFrame frame=new JFrame(title);
  return frame; }
\end{codesmall}
\vspace{5 pt}

\noindent At this point, the programmer accepts this suggestion, and uses the search result to replace the incomplete code fragment. Next, the programmer adds the following to the method:

\vspace{5 pt}
\begin{codesmall}
/**
 create button
*/
public ? ?(? a){
  __CODE_SEARCH__; }
\end{codesmall}
\vspace{5 pt}

\noindent
\textsc{Codec} now analyzes the entire class (including the new method just added), as well as the evidence present in the incomplete method (in this case, only the parameter name \texttt{a} is present) and returns the code among the top few results:

\begin{codesmall}
private JButton makeButton(String arg){
  JButton button=new JButton(arg);
  return button;
}
\end{codesmall}

\noindent Again, the programmer accepts this result and uses it to replace the fragment.
Now, the programmer adds the following method:

\begin{codesmall}
public void actionClose(JButton a, 
  JFrame f){
  __CODE_SEARCH__; }
\end{codesmall}

\noindent \textsc{Codec} uses all of the code so far to power the search, including the header for the incomplete method. The top result is:

\begin{codetiny}
 private void setCloseSurrogateButtonAction
   (JButton closeSurrogateButton,JFrame guiFrame){
   closeSurrogateButton.addActionListener 
     (new ActionListener() {
     public void actionPerformed(ActionEvent event){
       closeString = CLOSE_UI_EXIT_SURROGATE;
       guiFrame.dispatchEvent(new WindowEvent
         (guiFrame,WindowEvent.WINDOW_CLOSING));
       if (closingWindow) {surrogate.userExit=true;}
       else {closeString=
         CLOSE_UI_SURROGATE_KEEPS_RUNNING;}
     }});
 }
\end{codetiny}

\noindent Throughout the process, the programmer expended little effort to use the tool. No explicit queries were formulated beyond the development of the skeleton of the class.  \textsc{Codec} exclusively uses the context to anticipate what sort of example codes might be useful for the developer to consider. 

\section{Related Work}
Code search has been long-studied. Early works were information retrieval (IR) based. Classic methods include CodeBroker \cite{codebroker}, which assessed similarity using comments. 
Stratcona \cite{stratcona} used inheritance, method calls, and data types to compute similarity. XSnippet \cite{XSnippet} used parent classes and type information, and Bajracharya et al. \cite{Bajracharya} implement keyword-based API pattern search. 
More modern IR-based code search engines such as Koders \cite{Koders}, Krugle \cite{Krugle}, Codase \cite{Codase}, and Sourcerer \cite{sourcerer}
use text and graph-based search ranking. 

Another line of work powers code search using semantic or syntactic constraints. 
Prospector \cite{prospector} searches based on return types and types used in the code. JSearch \cite{JSearch}, 
Little et al. \cite{Little_ase07}, and PARSEWeb \cite{PARSEWeb} search using ASTs and API call sequences.
Reiss et al. \cite{Reiss_icse09} and CodeGenie \cite{CodeGenie} use 
test cases, contract specifications, and keywords from unfinished code to facilitate search. CodeHow \cite{CodeHow} uses keywords from natural language description and reinforces the search with an additional API understanding phase by mapping the keywords to the descriptions available in an online API library. Facoy \cite{facoy} proposes a code-to-code search methodology for detecting semantically similar code fragments by code alteration.

Modern code search tools use learned embeddings. 
Sachdev \cite{facebook_mapl18} suggests natural language based search using code embeddings and term frequency-inverse document frequency. Bajracharya \cite{Bajracharya} learns custom program and query embeddings mined from open-source data. \cite{doc_sim_icse16,deep_code_search} map natural language code descriptions and programs to a shared latent space. Lili \cite{Lili_AAAI16} propose a tree-based convolutional neural network to embed codes.  Chen \cite{ChenASE18} propose a variational-autoencoder-based architecture for code retrieval and summarization. Cambronero \cite{facebook_new_paper} does a comparison between the different neural code engines that use natural language for program search. A recent work by Wan \cite{wan2019multi} proposes using abstract syntax trees (ASTs) for a more accurate representation of programs, to facilitate natural language search. 

Recently there has also been interest in using deep learning-based techniques in other applications of software engineering. Neural-machine translation-based approaches have found application in detecting code clones in software repositories \cite{clone_det:codit, clone_det:tufano, clone_det:yu}. Our work has connections with recent attempts at using learning-based methods for program synthesis \cite{gulwani_synthesis}. The idea of using deep neural models for code completion in IDEs has also been popular recently. Pythia \cite{pythia} is an API recommendation engine that predicts likely API calls.  Pythia is integrated as part of Visual Studio for Python. Other works \cite{DeepCoder,parisotto_arxiv,robustfill} use ML to guide program synthesis.  \textsc{Bayou} \cite{bayou} synthesizes programs into a high-level representation; the \textsc{Sketch} language (see Figure \ref{fig:sketchlang}) was proposed in that paper and much of our statistical model was borrowed from Bayou. Program context has also been used to guide synthesis \cite{context_syn}.  
One of our contributions is bridging the gap between synthesis and search.

The idea of using different program components (return types, API call sequences, parent class information, and so on) as context for judging programmer intent have been widespread \cite{XSnippet, prospector, PARSEWeb, CodeGenie, stratcona}. Much recent work has applied deep learning for code search \cite{deep_code_search, facebook_mapl18, doc_sim_icse16}. However, this latter category of methods has generally been restricted to search based upon natural language specifications (for example, using the information contained in a JavaDoc to learn how to relate text to code).  To the best of our knowledge, our efforts are the first to use neural methods to power search using context.

%
%
%
%
%
%
%
%
%

\section{The \textsc{Codec} System}\label{sec:system}

\begin{figure*}[!t]
    \centering
    \includegraphics[width=1\textwidth]{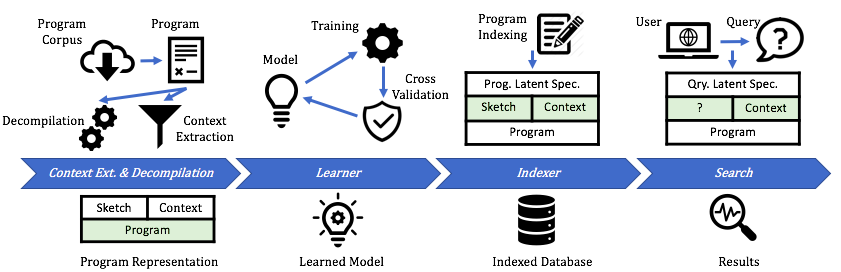}
    \caption{Schematic of the \textsc{Codec} system.}
    \label{fig:code_search_system}
\end{figure*}

In this section, we describe the design and implementation of  \textsc{Codec} at a high level.  A pictorial representation of the  \textsc{Codec} system is shown in Figure \ref{fig:code_search_system}. The system has five components. Note that while the current implementation of  \textsc{Codec} is specific to Java, extension to other programming languages is straightforward. 

\vspace{5 pt}
\noindent
\textbf{(1) Context extractor.} This component accepts a Java program, and uses the Eclipse compiler \cite{dom_driver} to parse it into an AST. From the program AST, each program fragment and surrounding context are extracted. Processing a Java program with the context extractor results in a set of $(\sX, \sProg)$ pairs---that is, a set of (context, code fragment) pairs.

\vspace{5 pt}
\noindent
\textbf{(2) Decompiler.} Modern Java is an exceedingly complex language, and many of the lower-level details associated with a Java code fragment are likely unimportant for deciding whether the code fragment answers a particular query. Thus,  \textsc{Codec} decompiles each code fragment $\sProg$ into a simpler programming language called \textsc{Sketch} that captures the essence of the Java fragment: API calls, types, general code ``shape'' (that is control flow and nesting) but ignores lower-level details such as variables and computation of expressions. The decompiler is realized by a function $\alpha$, such that the \textsc{Sketch} program $\sY = \alpha(\sProg)$.

\vspace{5 pt}
\noindent
\textbf{(3) Learner.} Given a code corpus to be indexed for search, all of the programs are fed into the context extractor, and the resulting code fragments are decompiled into \textsc{Sketch} codes.  A subset of the decompiled codes is used to create the training set $D_{trn} = \{\sX_i, \sY_i \}_{i = 1...n}$.  These pairs are then used to power a maximum likelihood estimation, where the goal is to choose the parameter set $\theta^{*}$ as
$$\theta^{*} = \textrm{argmax}_\theta \sum_i \log P(\sY_i | \sX_i, \theta)$$

\vspace{5 pt}
\noindent
\textbf{(4) Indexer.} Once the parameter set $\theta$ has been learned, each \textsc{Sketch} code $\sY$ to be indexed is then transformed into an intermediate representation $\sY' = h(\sY, \theta)$ that makes it very fast to compute $P(\sY | \sX, \theta)$ for a context query $\sX$. The set $D_{srch} = \{\sY'_i, \sProg_i \}_{i = 1...n}$ where $\sY'_i = h(\alpha(\sProg_i))$ is then distributed across the set of servers used to power the search.

\vspace{5 pt}
\noindent
\textbf{(5) Distributed search engine.} Finally, at query time, query context $\sX_q$ is automatically extracted from a partially-completed program, and $D_{srch}$ is processed to compute the top $K$ $(\sY'_i, \sProg_i )$ pairs from  $D_{srch}$ that maximize $P(\sProg_i | \sX_q, \theta)$, which are presented to the user.

\vspace{5 pt}
\noindent
In the next few sections of the paper, we describe a few of these components in more detail, followed by an in-depth description of the statistical model $P(\sProg | \sX, \theta)$ used to power the search, as well as how this model is learned from $D_{trn}$.

\section{Context Extraction and Decompilation}

In this section, we describe program context extraction and decompilation in a bit more detail.

\subsection{Context Extraction} \label{sec:context_extraction}

Given a Java program, the first step is to parse the program, and extract the various \emph{evidences} that serve as the context for all of the code fragments that will be extracted from the code.  In \textsc{Codec}, some evidences are extracted on a per-class basis, so that all of the fragments from the same class will share the same evidence. The class-wide evidences extracted are:

\vspace{5 pt}
\noindent 
(1) \emph{Class Name}; name of the class split using camel-case; 
each is encoded with a one layer GRU-RNN network.\\
(2) \emph{Class types}; types of the instance variables in the same class as the query method; 
each is encoded with a one-hidden-layer fully-connected network.\\
(3) \emph{Surrounding Methods}; methods within the same class; each consisting of: \\
(a) \emph{Return type}; encoded with a one-hidden-layer network. \\ 
(b) \emph{Input parameter list}; sequence of (formal parameter, variable name) pairs encoded using a two-layer GRU-RNN, where the variable names are split using camel-case, encoded using a GRU-RNN, and concatenated with each formal parameter. \\
(c) \emph{Method name}; name of the method split using camel-case; each is encoded with a one layer GRU-RNN network. \\
(d) \emph{API call sequences}; API call sequences are extracted from methods within the same class; encoded with a one layer GRU-RNN network.

\vspace{5 pt}
\noindent 
We also use four types of evidence from the header of the missing method (if available):

\vspace{5 pt}
\noindent
(1) \emph{JavaDoc};
English text of JavaDoc associated with the method, lemmatized and stop words removed.  Encoded using a  
bidirectional GRU-RNN.  \\
(2) \emph{Method name} for method containing the missing code; split using camel-case and encoded with a single-layer GRU-RNN. \\ 
(3) \emph{Return type} of method with the missing code encoded with a one hidden-layer network. \\
(4) \emph{Input parameter list}, of method with missing code; including formal parameter type and name, split using camel-case, encoded similarly to the input parameters from surrounding methods.

\vspace{5 pt}
\noindent 
Finally, we use four types of evidence from \emph{within} the (partial) code that is being searched for. These may be available if a fragment of the code has already been written. They are: 

\vspace{5 pt}
\noindent
(1) \emph{API calls}; these are the calls in the code; each is encoded with a one-layer network. \\
(2) \emph{API call sequences}; these are extracted via symbolic execution of the code; each is encoded using a GRU-RNN. \\
(3) \emph{Types}; these are the API types in the code; each is encoded with a one-layer network. \\
(4) \emph{Keywords}; English-like words extracted from fully qualified name of the classes inside the method body, combined with English words appearing in types and API calls; each is encoded with a one-layer network.

\begin{figure}
$
\begin{array}{lll}
\Sy & ::= & \Sy_{api};\Sy_{ret};\Sy_{fp} \\
\Sy_{ret} & ::= & \tau_{r}\\
\Sy_{fp} & ::= & (\tau_{fp_0},\dots,\tau_{fp_n})\\
\Sy_{api} & ::= & \skipc ~|~ \callc~\id{Cexp} ~|~ \Sy_1; \Sy_2 ~|~\smallskip \\
& & \ifc~\id{Cseq}~\thenc~\Sy_1~\elsec~\Sy_2 ~|~\smallskip \\
& &  \whilec~\id{Cseq}~\doc~\Sy_1 ~|~ \tryc~\Sy_2~\id{Catch}
                \\
\id{Cexp} & ::= & \tau_{a_0}.\alpha(\tau_{a_1},\dots, \tau_{a_k}) \smallskip \\
\id{Cseq} & ::= & \textrm{List of}~\id{Cexp} \smallskip \\
\id{Catch} & ::= & \catchc (\tau_{a_1})~\Sy_1~\dots~\catchc(\tau_{a_k})~\Sy_k
\end{array}
$
\caption{Grammar for the \textsc{Sketch} language. $\tau$ indicates a Java type, $\alpha$ is a method call name. \textsc{Sketch} extends the \textsc{Bayou} sketching language [32]}
\label{fig:sketchlang}
\end{figure}

\subsection{Decompilation} \label{sec:decompilation}

As described previously, we believe that it is problematic to search for a code fragment in a complicated language such as Java directly. Every high-level language likely contains details (such as arithmetic operations) that are of little use during search, and likely make it difficult to learn how to relate queries with search results, obscuring the important facets of the code.  A complicated and mature language such as Java is especially problematic. Consider the \textsc{Bayou} program synthesis system \cite{bayou}. Using a neural network to synthesize into a sketching language (and then using classical, AI-style search to complete the program) Bayou showed around 50\% accuracy (in terms of being able to reproduce the ``correct'' result in the top-10 programs synthesized), whereas a version of Bayou that synthesized directly into a subset of Java was able to achieve less than 10\% in terms of top-10 accuracy.

Hence, given a class, once we extract a code fragment $\sProg$ from the class, we \emph{decompile} $\sProg$ into a simplified representation $\sY = \alpha (\sProg)$.  This representation is designed to retain the facets of the code that are likely to be important to a user during search: API calls and basic control flow, but ignores the rest.  The grammar for the abstraction language \textsc{Sketch} is given above in Figure ~\ref{fig:sketchlang}.  Translating from a parsed Java AST to a \textsc{Sketch} AST is straightforward.

For an example of this, consider the following method:

\begin{codesmall}
void read(File file) {
  FileReader fr1;
  BufferedReader br1;
  fr1 = new FileReader(file);
  br1 = new BufferedReader(fr1);
  while ((br1.readLine()) != null) {}
  return; }
\end{codesmall}

\noindent
This is decompiled into the following:

\begin{codesmall}
FileReader.FileReader (File)
BufferedReader.BufferedReader (FileReader)
while
  BufferedReader.readLine ()
do
  skip
\end{codesmall}

\noindent
Note the \texttt{skip}, which reflects the fact that the \texttt{while} loop has no body.

\section{Statistical Model}

\label{sec:forward_model}

At the heart of the \textsc{Codec} system is the statistical model powering the search, embodied by the distribution function $P(\sProg|\sX, \theta)$.  During search, a set of evidences is extracted to represent the program context $\sX$, and then the few database fragments that maximize the value of this function are selected.

\begin{figure}[t]
	\centering
		\centering
		\begin{tikzpicture}
		\tikzset{vertex/.style = {shape=circle,draw}}
		\tikzset{edge/.style = {->,> = latex'}}
		
		\node[vertex] (x) at  (-1.5,0) {$~X~$};
		\node[vertex] (z) at  (0,1) {$~Z~$};
		\node[vertex] (y) at  (1.5,0) {$~Y~$};
		\node[vertex] (p) at  (3.5,0) {$~Prog~$};
		\draw[edge] (z) to (x);
		\draw[edge] (z) to (y);
		\draw[edge] (y) to (p);
		\end{tikzpicture}
		
	\caption{Bayes net for $X$, $Y$, $Z$ and $Prog$.}
	\vspace{-7 pt}
	\label{fig:Bayes}
\end{figure}
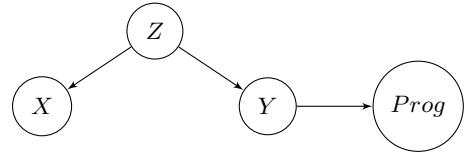

There are many possible choices for the model
$P(\sProg|\sX, \theta)$. 
We begin with the simple statistical model pictured in Figure \ref{fig:Bayes}, borrowed from conditional program generation \cite{bayou} .
Crucially, we assume a latent specification $Z$ for the missing code fragment as well as the surrounding context.  This specification captures the programmer intent, and conditioned upon $Z$, both the program context $X$ as well as the sketch $Y$ and the program fragment itself is generated.

As intimated in the introduction, there is a key benefit to using such a model to power code search. During search, the latent variable $Z$ provides similar functionality to the latent-space embedding used in traditional, embedding-based methods \cite{semantic_hashing}. However, there is a key difference.  As $Z$ is a true random variable, it has no single value. A learner, given a large number of $(\sX, \sProg)$ pairs from which to learn $P(\sProg|\sX, \theta)$, must learn to accurately generate $\sProg$, despite of the uncertainty in $\sZ$ embodied by the distribution $P(Z)$.  This may alleviate some of the problems with over-fitting one might except when using a more traditional neural encoding.

As depicted in Figure \ref{fig:Bayes}, $Z$ is generated first, and based upon the programmer intent captured by $Z$, the evidences $X$ in the surrounding context are generated, as well as the sketch $Y$. Once the sketch is generated, the program $\sProg$ is generated based on the sketch.  Thus, the joint distribution $P(X, Y, Z, Prog | \theta)$ factorizes as $P(\sX, \sY, \sZ, \sProg | \theta) =$ $$P(\sZ) P(\sX | \sZ) P(\sY |  \sZ) P(\sProg | \sY).$$
\noindent (Note that we drop the parameter $\theta$ from each distribution function for simplicity).

\section{Search Under the Model}

\subsection{Applying the Model for Search}

During search, we are given a context $\sX$, and we wish to choose a database code fragment $\sProg_i$ to maximize 
$P(\sProg_i | \sX, \M) =$ $$\int_\sY P(\sProg_i | \sY, \M) \int_{\sZ} P(\sY | \sZ, \M) P(\sZ | \sX, \M) d\sZ d\sY $$
\noindent This looks difficult. However, we can simplify this expression by assuming that no code $\sProg_i$ is associated with more than one sketch; this is a fairly weak assumption, and is implied by the fact that we can decompile each code into its unique sketch using the decompilation function $\alpha$.  Given this assumption, the distribution function $P(\sProg_i | \sY, \M)$ gives non-zero likelihood for only one $\sY$ value for a given program $\sProg_i$.  Hence, if we let $\sY_i = \alpha(\sProg_i)$, $P(\sProg_i | \sX, \M)$ can be re-written as: 
$$P(\sProg_i | \sX, \M) = P(\sProg_i | \sY_i, \M) \int_{\sZ} P(\sY_i | \sZ, \M) P(\sZ | \sX, \M) d\sZ $$
\noindent In our implementation of \textsc{Codec} we simplify this further by assuming that the process of creating a code from a sketch is deterministic, and that $P(\sProg | \sY, \M) \neq 0$ if and only if $\sY = \alpha(\sProg_i)$, so that:
$$P(\sProg_i | \sX, \M) = P(\sY_i | \sX, \M) = \int_{\sZ} P(\sY_i | \sZ, \M) P(\sZ | \sX, \M) d\sZ $$
Though this assumption is not necessary, it seems to give good results, and it means there is no need to define $P(Prog | \sY, \M)$.   

\subsection{Making Search Fast} \label{sec:search-fast}

$P(\sY_i | \sX, \M)$ will need to be evaluated for millions or billions of $Y_i$ values stored in a database, in response to a query $\sX$.  For this reason, evaluating $P(\sY_i | \sX, \M)$ needs to be very, very fast.  Without a careful choice of the various distribution functions to allow for a fast, closed-form evaluation of $P(\sY_i | \sX, \M)$, search will not be practical.

To ensure that we are able to have a closed-form evaluation of this function, we begin by expanding the function using Bayes' rule: 
\begin{equation} \nonumber
\begin{aligned}
\int_{\sZ} P (\sY | \sZ, \M) & P(\sZ | \sX, \M) d\sZ \approx \\
&P(\sY | \M) \times \int_{\sZ} \frac{   P (\sZ | \sX, \M) P (\sZ | \sY, \M)}{P(\sZ | \M)}  d\sZ
\end{aligned}
\end{equation}

It can be shown that as long as all of the distribution functions within the integral ($P (\sZ | \sX, \M)$, $P (\sZ | \sY, \M)$, and $P(\sZ | \M)$) are multivariate Gaussian, this can be integrated analytically, with very little computational cost.\footnote{Intuitively, this is not surprising, as the Gaussian distribution is simply an exponentiated quadratic function, and every exponentiated quadratic function is a Gaussian distribution, up to a constant factor. Hence, multiplying and dividing Gaussian distribution functions results in an exponentiated quadratic function, which is then also a Gaussian distribution function, up to a constant factor. Integrating over any distribution function results in a value of one, leaving only the constant factor. If we can compute that constant analytically, there is no need to integrate.}
For example, assume $Z$ is scalar-valued,\footnote{An extension to multiple dimensions is straightforward and a full derivation with multivariate extension is available in the Appendix section.} and assume the mean and variance of $P (\sZ | \sX, \M)$ evaluate to $\mu_{\sX}$
and $\sigma_{\sX}^2$, respectively. 
Also assume that the mean and variance of $P (\sZ | \sY, \M)$ evaluate
to $\mu_{\sY}$
and $\sigma_{\sY}^2$, respectively.  These can be computed offline, during database preparation, and stored in the
database along with the programs to be searched.

\vspace{6 pt}
Let $a_{\sX} = -(2\sigma_{\sX}^2)^{-1}$ and define 
$a_{\sY}$ similarly.  Likewise, let $b_{\sX} = \mu_{\sX}\sigma_{\sX}^{-2}$ and define $b_{\sY}$ similarly. Finally, we let $P(\sZ | \M)$ be unit multivariate Normal.
Then we have:
\begin{equation} \nonumber
\begin{aligned}
\log P(\sY | \sX, \M) =& \log P(\sY | \M) + \frac{1}{2}\log \left(\frac{-2a_{\sX}a_{\sY}}
{a_{\sX} + a_{\sY} + 1/2} \right)  \\
&+ \frac{b_{\sX}^2}{4a_{\sX}} + \frac{b_{\sY}^2}{4a_{\sY}} -
\frac{(b_{\sX} + b_{\sY})^2}{4(a_{\sX} + a_{\sY} + 1/2)}
\end{aligned}
\end{equation}
\vspace{2 pt}
Except for the quantity $\log P(\sY | \M)$, this is all trivial to evaluate quickly, at search time, and so it is computationally efficient to check $\log P(\sY_i | \sX, \M)$ for each $\sY_i$ in the database as long as we have pre-computed and stored three values:  

\begin{enumerate}[leftmargin=*]

\item The term
$\log P(\sY_i | \M)$, which measures the bias towards returning a particular program, can be computed offline via
Monte Carlo integration over the latent variable $\sZ$, using $P (\sZ | \sY_i, \M)$ as a 
proposal distribution for importance sampling \cite{mcmc}.

\item The mean and variance of $P (\sZ | \sY_i, \M)$.  In the univariate case, these will be scalar values, and in the multivariate case these will be vector-valued (assuming a diagonal covariance matrix for $P (\sZ | \sY_i, \M)$, as we will assume subsequently).

\end{enumerate}

\noindent Together, these values constitute $\sY'_i$. In parallel across multiple compute nodes, \textsc{Codec} stores the set of programs to search $D_{srch} = \{\sY'_i, \sProg_i \}_{i = 1...n}$. In response to a query context $\sX$, the few $\sProg_i$ values for which $Y_i = \alpha(\sProg_i)$  maximizes the $\log P(\sY | \sX, \M)$ value are returned to the user.

\section{Distribution Families Used} \label{sec:encoding}

So far, we have not committed to any particular distribution functions, other than stating that our basic statistical model as shown in Figure \ref{fig:Bayes}, and stating that for
practical reasons---we want the per-program computation time to be tiny during search---we will desire each of
$P(Z| \M)$, $P(Z|\sX, \M)$ and $P(Z|\sY, \M)$ to be multivariate Gaussian. In this section, we discuss each of these distributions in more detail. 

\vspace{10 pt}
\noindent
\textbf{The prior on $Z$:} $P(Z | \M)$. Ensuring that $P(Z| \M)$ is multivariate Gaussian is easy; since $Z$ is a latent variable, we simply define $Z \sim$ 
Normal $(\vec{0}, \textbf{I})$.  

\vspace{10 pt}
\noindent
\textbf{$Z$ conditioned on the sketch}: $P(Z|\sY, \M)$. Unfortunately, marrying the natural set of conditional dependencies depicted in Figure~\ref{fig:Bayes}
with the desire for computational efficiency is not easy for the other distribution functions. In particular,
ensuring Normality for $P(Z|\sY, \M)$ is, practically speaking, not possible.  Since a given $\sY$ is a complex,
tree-valued object (a set of recursive rule firings in a grammar) we wish to use a state-of-the-art neural tree decoder to realize $P(Y|\sZ, \M)$. Several are available \cite{manning_decoder1, manning_decoder2, tree_decoder3};  we
use a top-down tree LSTM \cite{top-down-tree-lstm} to realize
$P(Y|\sZ, \M)$.  
However, using such a decoder almost assuredly means that $P(Z|\sY, \M)$ is \emph{not} Gaussian.  As we describe in the
next section of the paper, we address this by using variational methods \cite{vae} to simultaneously learn a Gaussian
approximation $Q(Z|\sY, \M)$ for $P(Z|\sY, \M)$, and to force $P(Z|\sY, \M)$ to be approximately Gaussian.
Then $Q(Z|\sY, \M)$ can then be used in place of $P(Z|\sY, \M)$ to ensure fast search.
We call $Q(Z|\sY, \M)$ a ``reverse encoder'' for a decompiled program $\sY$, as it reverses the generative process
to encode the program.
In our implementation, $Q(Z|\sY, \M)$ is realized as a neural tree encoder \cite{top-down-tree-lstm}, that is
used to encode $\sY$ into a mean vector
and covariance matrix of a Gaussian distribution.

\vspace{10 pt}
\noindent
\textbf{$Z$ conditioned on the context}: $P(Z|\sX, \M)$. This situation is a bit different.
We could also use variational methods to allow for a Gaussian approximation to $P(Z|\sX, \M)$, but instead we borrow the
formulation from \cite{bayou} to ensure that $P(Z|\sX, \M)$ \emph{is} Gaussian, at least under certain restricted 
circumstances.  

\vspace{2 pt}
Specifically, we assume that the context $\sX$ is partitioned into a set of different sets of evidences, according to the type
of evidence. Let $\sX_{j,k}$ refer to the $k$th instance of the $j$th type of evidence in $\sX$.  Various possible
types of evidence are discussed in Section~\ref{sec:context_extraction}, and include: class variable types, other method
signatures, documentation, etc.  Assume a neural function $f_{j, \M}$ for the $j$th evidence type that maps some
representation of the evidence to some location in $\mathbb{R}^m$. Then
let:
\begin{equation} \nonumber
P(\sX | \sZ, \M) = \prod\limits_{j,k} \textrm{Normal} \left(f_{j, \M}(\sX_{j,k}) | 
		\sZ, \textbf{I} \sigma^2_j\right) 
\end{equation}
\noindent Effectively, we assume that the encoded location of each piece of evidence has been sampled from a
normal distribution with mean $\sZ$.  Each evidence is sampled using a different variance.  Higher variance corresponds
to an evidence that is less closely related with the functionality of the code fragment being searched for.

\vspace{3 pt}
If each mapping function is one-to-one and onto, then from Normal-Normal conjugacy, it follows that \cite{bayou}:
\begin{equation} \nonumber
P(\sZ | \sX, \M) = \mathcal{N} \left(\sZ | \frac{\sum\limits_{j,k} \sigma^{-2}_j f_{j, \M} (\sX_{j,k})}{1 + \sum\limits_j |\sX_{j}| \sigma^{-2}_j}, \frac{1}{1 + \sum\limits_j |\sX_{j}| \sigma^{-2}_j}\textbf{I}\right)
\end{equation}

\noindent Here, $|\sX_{j}|$ refers to the size of the $j$th subset of evidence. \\

Note that Normal-Normal conjugacy will not hold if some mapping function is not one-to-one and onto.  
In practice, this will \emph{not} hold. 
As an example, we may employ a bidirectional RNN \cite{bi-rnn} to encode English text in a
JavaDoc comment into $\mathbb{R}^m$; such a function will not be one-to-one and onto.  Still, in practice
things seem to work well, and intuitively the fact that the function is not one-to-one and onto
may only be a problem if different evidences tend to be mapped to the
same point in $\mathbb{R}^m$, which seems not to happen in practice.
As intimated above, an alternative is to instead 
resort to a variational approximation for $P(\sZ | \sX, \M)$ as well as $P(\sZ | \sY, \M)$, at the cost of
making the learning problem somewhat more complex.

\section{Training the Model} \label{sec:training}

We have two goals.  First, we wish to make sure that the reverse encoder $Q(Z|\sY)$ is a reasonable proxy for $P(Z|\sY)$.  Second, we wish to ensure that the log-likelihood of our data set is maximized and that $\sum_i \log P(\sY_i | \sX_i, \theta)$ has a large value.  In the remainder of this section, we again drop the set of model parameters $\M$ when convenient to simplify the presentation.

To begin, we note that we want the so-called ``reverse encoder'' $Q(Z|\sY)$ to closely match the true posterior $P(Z|\sY)$ for $\sY \sim P(Y)$.  We can ensure this by minimizing the KL divergence between them. We begin our derivation of the learning problem by expanding this KL divergence. For a given $\sY$:
\begin{equation} \nonumber
\begin{split}
D_{KL}&\big(Q(Z|\sY)\|P(Z|\sY)\big) \\
&=\int_Z Q (\sZ|\sY) \big[\log Q(\sZ|\sY) - \log P(\sZ|\sY)\big]d\sZ \\
\end{split}
\end{equation}
\noindent Note that $P(Z|\sY)$ cannot be evaluated directly, as it is not one of the three distributions defined in the previous section. Hence, we expand it using Bayes' Rule: 
\begin{equation} \nonumber
\begin{split}
&D_{KL}\big(Q(Z|\sY)\|P(Z|\sY)\big) \\
&=\int_Z Q(\sZ|\sY) \times \big[\log Q(\sZ|\sY) - \log P(Y|\sZ) \\
& \qquad \qquad \qquad \quad - \log P(\sZ) + \log P(\sY) \big] d\sZ  \\
&=\log P(\sY) + \int_Z Q(\sZ|\sY) \times \big[ \log Q(\sZ|\sY) \\
& \qquad \qquad \qquad \quad - \log P(\sY|\sZ)  - \log P(\sZ) \big] d\sZ  \\
\end{split}
\end{equation}
\noindent Expanding this further we have:
\begin{equation} \nonumber
\begin{split}
&D_{KL} \big(Q(Z|\sY)\|P(Z|\sY) \big) \\
&= \log P(\sY) +\int_\sZ Q(\sZ|\sY) \times \big[ \log Q(\sZ|\sY) - \log P(\sZ|\sX) \big] \\
& +\int_\sZ Q(\sZ|\sY) {\times} \big[\log P(\sZ|\sX) - \log P(\sY|\sZ)  - \log P(\sZ) \big] d\sZ  \\
\end{split}
\end{equation}
\begin{equation} \nonumber
\begin{split}
&= \log P(\sY) + D_{KL}(Q(Z|\sY)\|P(Z|\sX)) \\
& + \int_\sZ Q(Z|\sY) \big[\log P(Z|\sX) - \log P(Y|\sZ)  - \log P(Z) \big] d\sZ  \\
&= \log P(\sY) + D_{KL}(Q(Z|\sY)\|P(Z|\sX)) + \int_\sZ P(\sZ|\sX) \times  \\
& \qquad \frac{Q(\sZ|\sY)}{P(\sZ|\sX)} {\times} \big[ \log P(\sZ|\sX) - \log P(\sY|\sZ) - \log P(\sZ) \big] d\sZ  
\end{split}
\end{equation}
\noindent Assume for a moment that $Q(Z|\sY) \approx P(Z|\sX)$ for $(\sX, \sY) \sim P(X, Y)$ so their ratio is 1.  While this is by no means guaranteed, we will reconsider this assumption later. Then we have:
\begin{equation} \nonumber
\begin{split}
D_{KL}&\big(Q(Z|\sY)\|P(Z|\sY)\big) \approx \\
& \log P(\sY) + D_{KL}\big(Q(Z|\sY)\|P(Z|\sX)\big) \\
& + \int_\sZ P(\sZ|\sX)  \times \big[\log P(\sZ|\sX) - \log P(\sZ)\big] d\sZ \\
& - \int_\sZ P(\sZ|\sX) \times  \log P(\sY|\sZ) d\sZ
\end{split}
\end{equation}
\noindent This can be rewritten as,
\begin{equation}  \nonumber
\begin{split}
D_{KL}&\big(Q(Z|\sY)\|P(Z|\sY)\big)  \approx \\
&\log P(\sY) + D_{KL}\big(Q(Z|\sY)\|P(Z|\sX)\big) \\
& +D_{KL} \big(P(Z|\sX)\|P(Z)\big)  -\int_\sZ P(\sZ|\sX) \log P(\sY|\sZ) d\sZ  \\
\end{split}
\end{equation}
\noindent And so, 
\begin{equation}  \nonumber
\begin{split}
\log &P(\sY) - D_{KL}\big(Q(Z|\sY)\|P(Z|\sY)\big) \approx \\
&-D_{KL}\big(Q(Z|\sY)\|P(Z|\sX)\big) \\
& -D_{KL} \big(P(Z|\sX)\|P(Z)\big)  +\int_\sZ P(\sZ|\sX) \log P(\sY|\sZ) d\sZ   \\
\end{split}
\end{equation}

\noindent This implies that if we maximize the
expected value of the RHS of the above approximation with respect to $(\sX, \sY) \sim D_{trn}$ we will 
simultaneously perform a maximum likelihood estimation (maximizing the data log-likelihood $\log P(Y)$) and maximize the quality of the reverse encoder $Q(Z|\sY)$ by making it a good approximation for $P(Z|\sY)$.  Then, 
in the end, to learn the model, we choose $\theta$ so as to maximize the following:
\begin{equation}  \nonumber
\begin{split}
\mathrm{E}_{(\sX, \sY) \sim D_{trn}} \bigg[
&-D_{KL}\big(Q(Z|\sY)\|P(Z|\sX)\big) \\
& -D_{KL} \big(P(Z|\sX)\|P(Z)\big) \\ 
&+\int_\sZ P(\sZ|\sX) \log P(\sY|\sZ) d\sZ  \bigg]
\end{split}
\end{equation}

This is easily possible via gradient descent.  We sample $(\sX, \sY)$ pairs from $D_{trn}$, and for each pair, take a gradient step to minimize the value of the expression.  Fortunately, since each $Q(Z|\sY)$, $P(Z|\sX)$, and $P(Z)$ is multivariate Gaussian, there is a closed form for the pairwise KL divergence between them for which the gradient is easily computed using a platform such as TensorFlow.  

\vspace{5 pt} 
One more complicated issue is to compute the gradient of the integral $\int_\sZ P(\sZ|\sX) \log P(\sY|\sZ) d\sZ$.  Note that this can be re-written as
$\mathrm{E}_{(\sZ) \sim P(Z|\sX)} \log P(\sY|\sZ)$. Maximization of this quantity is 
amenable to the standard ``reparameterization trick'' \cite{vae} used when training variational 
autoencoders.  That is, we
may sample $\sZ$ from a standard Normal distribution, and then push
the transformations represented by $P (Z | \sX, \M)$ and $Q (Z | \sY, \M)$ into the quantity we are taking the
expectation of, in order to back-propagate through the transformations.

\vspace{5 pt} 
Finally, we re-visit our assumption that $\frac{Q(\sZ|\sY)}{P(\sZ|\sX)} \approx 1$. While not guaranteed, the argument for the validity of this simplifying assumption rests on the fact that the resulting maximization problem explicitly attempts to minimize the KL divergence term in our derivation, $D_{KL}\big(Q(Z|\sY)\|P(Z|\sX)\big)$ for $(\sX, \sY) \sim D_{trn}$.  As this divergence is minimized during learning, the approximation will become increasingly valid.

\vspace{5 pt} 
\noindent
\textbf{Relationship to variational autoencoders.}
There is a resemblance between the material in this Section and the methods used to train variational autoencoders (VAEs) \cite{vae}. However, there are key differences. Given a data \textsf{Y}, a VAE is meant to learn a model of the form $P(Y = \textsf{Y}) = \int P(\textsf{Z}) P(\textsf{Y}|\textsf{Z}) d \textsf{Z}$, whereas our goal is to learn a conditional model of the form $P(Y = \textsf{Y} | \textsf{X}) = \int P(\textsf{Z}|\textsf{X}) P(\textsf{Y}|\textsf{Z}) d \textsf{Z}$.   In the case of a VAE, a variational distribution $Q(Z |  \textsf{Y})$ is used to approximate $P(Z |  \textsf{Y})$ to make training possible.  In our case, because of the availability of the evidence set $\textsf{X}$ (as in conditional program generation \cite{bayou}) we do not need this approximation; it would be possible to learn $P(Y = \textsf{Y} | \textsf{X})$ directly without any variational approximation, using a bound based on Jensen's inequality. However, we also have to simultaneously learn an approximation $Q(Z |  \textsf{Y})$ for $P(Z |  \textsf{Y})$ to use during code search.  This results in a learning problem formulation that differs in important ways from VAEs, conditional VAEs \cite{conditional_vae}, and 
conditional program generation.

\section{Evaluation}
\begin{table*}[ht]
	\centering
	\caption{Prediction accuracy comparison from program context.}
	\vspace{-3 pt}
	\begin{tabular}{|l|c|c|c|c||c|c|c|c|}
	    \hline
	     & \multicolumn{4}{c||}{\textbf{SuccessRate@1}} & \multicolumn{4}{c|}{\textbf{SuccessRate@10}} \\
		\hline
	     & API Match & Seq Match & Sk. Match & Exact Match & API Match & Seq Match & Sk. Match & Exact Match \\
		\hline
		\hline
		CodeHow & 0.03 & 0.02 & 0.02 & 0.00 & 0.12 & 0.09 & 0.09 & 0.06 \\
		\hline
        Deep-Code & 0.04 & 0.02  & 0.00 & 0.00  & 0.11 & 0.06 & 0.03 & 0.02 \\
		\hline
		Non-Prob & 0.01 & 0.01 & 0.01 & 0.00 & 0.02 & 0.01 & 0.01 & 0.00  \\
		\hline
		\textsc{Codec} & \textbf{0.27} & \textbf{0.24} & \textbf{0.23} & \textbf{0.11} & \textbf{0.35} & \textbf{0.32} & \textbf{0.32} & \textbf{0.15} \\
		\hline
		\hline
		 & \multicolumn{4}{c||}{\textbf{Precision@10}} & \multicolumn{4}{c|}{\textbf{MRR}} \\
		\hline
	     & API Match & Seq Match & Sk. Match & Exact Match & API Match & Seq Match & Sk. Match & Exact Match \\
		\hline
		\hline
		CodeHow & 0.04 & 0.03 & 0.03 & 0.01 & 0.13 & 0.12 & 0.12 & 0.10 \\
		\hline
        Deep-Code & 0.04 & 0.02 & 0.00 & 0.00 & 0.15 & 0.12  & 0.10 & 0.10 \\
		\hline
		Non-Prob & 0.01 & 0.00 & 0.00 & 0.001 & 0.11 & 0.11 & 0.11 & 0.09    \\
		\hline
		\textsc{Codec} & \textbf{0.26} & \textbf{0.24} & \textbf{0.22} & \textbf{0.10} & \textbf{0.35} & \textbf{0.33} & \textbf{0.32} & \textbf{0.20} \\
		\hline
	\end{tabular}
	\label{comparison_matrix}
\end{table*}

There are three parts to our experimental study of \textsc{Codec}. 

In the first, we perform a quantitative study where “holes” are created in a large number of real-life programs, downloaded from GitHub, by removing a method body from each program. The method bodies are then mixed into a large database of method bodies, and we measure  \textsc{Codec}’s ability to retrieve the correct method body from the database

In the second, we perform a qualitative user study where we ask 13 programmers to rate the quality of the results returned by \textsc{Codec} and a number of competitive methods.

In the third, we examine the runtime efficiency of our so-called ``reverse encoder,'' that enables a fast, analytic approximation to the likelihood that each program was generated by the search context.

\subsection{Quantitative Study}

\subsubsection{Experimental Setup}

\vspace{5 pt} \noindent 
\textbf{Data used.} We collect all the public, licensed projects available in Github \cite{github}.  This is a total of 8.71M java files. We use the Eclipse Java DOM Driver \cite{dom_driver} to extract abstract syntax trees for a total of 27.9M methods with at-least one API call to the Java JDK. Out of these, we trained our statistical model (as well as the competitive models described below) on 2.32M randomly-sampled Java files, amounting to 6.7M methods. 21M Java methods were extracted from the remaining 6.39M files, and the methods extracted were indexed using the learned model. 

\vspace{5 pt} \noindent 
 \textbf{Model training details.} The latent space occupied by \textit{Z} in our implementation is 256 dimensions. We have used 256 units in each of our neural architectures and a single hidden layer for our tree encoder and decoder. We used a batch-size of 128 methods during training and a learning rate 0.0001 for the Adam gradient descent algorithm \cite{adam}. Our deep  model was trained on top of Tensorflow \cite{tensorflow} using an Amazon EC2 \texttt{p2.xlarge} machine powered with an NVIDIA K80 GPU. Training required 200 hours. 

\vspace{5 pt} \noindent 
\textbf{Competitive methods.}  We compare \textsc{Codec} with three baseline methods.
The first two are CodeHow \cite{CodeHow} and Deep-Code search \cite{deep_code_search}. Both of these methods were developed to support code search using natural language.  CodeHow finds programs based on keyword matching, using API understanding to reformulate the query for higher accuracy.  We modify CodeHow to use keywords from method headers (types, formal parameter names, method names) along with JavaDoc comments. Deep-Code search encodes the JavaDoc and the program (represented as a triplet of method name, sequence of API calls and keywords) into a shared latent space and attempts to minimize the cosine distance between them. We also implement a non-probabilistic version of \textsc{Codec} that uses the same encoders for various evidences as \textsc{Codec}, encoding the entire context as a weighted average of the various evidences. This non-probabilistic version uses the same neural architecture as \textsc{Codec}'s reverse encoder to encode the sketch into the latent space, and attempts to maximize the cosine similarity between the encoded context and the encoded sketch.

\vspace{5 pt} \noindent 
\textbf{Retrieval task.} We test each method using a set of 100 retrieval tasks.  To construct a task, we randomly select a Java file having at least two method bodies from among the 6.4M Java files not used for training.
The files selected represent a wide variety of Java applications, from computer networking applications to MySQL database application development to simple file I/O. We remove a random method body containing JavaDoc documentation from the file, and use the context in the remainder of the file to power search using the four different search techniques.  The search is considered to be accurate if one or more method bodies that is ``equivalent'' to the removed-and-searched-for method body are among the top results returned.

\vspace{5 pt} \noindent 
\textbf{Measuring ``equivalence.''}  Defining the notion of two codes being equivalent to one another is not straightforward. Determining if two codes produce the same output on all inputs is, in general, undecidable, and a real-life, GitHub-derived Java corpus presents many challenges.  For example, a popular code may be replicated any times during its lifetime on GitHub.  If the retrieval task returns an older version of the correct method, it is unclear whether this is ``equivalent.''  

In the end, we came up with four different definitions of method equivalence, each of which we examine experimentally: (1) \textit{API match}; two codes are  equivalent if two use the same set of JDK API calls. (2) \textit{Sequence match}; two codes are equivalent if the sets of all possible sequences of API calls, extracted using symbolic execution, are the same. (3) \textit{Sketch match}; two codes are equivalent if decompilation into a \textsc{Sketch} program (see Section \ref{sec:decompilation}), results in the same code.
(4) \textit{Exact match}; two programs are considered to be equivalent if the Java parse tree for the entire method body matches exactly.

\begin{table}[!t]
	\centering
	\caption{Search problems considered in user study.}
	\vspace{- 3 pt}
	\small
	\begin{tabular}{|l|l|l|}
		\hline
		\textbf{Id} & \textbf{Program Class} & \textbf{Programming Task}\\
		\hline
		\hline
		\textbf{1} & Socket & Send data using Client \\
		\hline
		\textbf{2} & Crypto & Encrypt data using MD5 hash \\
		\hline
		\textbf{3} & FileUtils & Copy a file to a different location \\
		\hline
		\textbf{4} & Generic list & Remove item from list \\
		\hline
		\textbf{5} & GUI-Swing & Add closing button to a frame\\
		\hline
		\textbf{6} & Conversion & Convert a list of string to HashMap \\
		\hline
		\textbf{7} & IO & Write data using \texttt{OutputStream} \\
		\hline
		\textbf{8} & String Operations & Check if \texttt{String} is palindrome \\
		\hline
	    \textbf{9} & Parser & Parse JSON \texttt{String}; put into hash \\
		\hline
		\textbf{10} & Peeking Iterator & Advance iterator if next exists  \\
		\hline
		\textbf{11} & SQL & Execute a select statement \\
		\hline
		\textbf{12} & Stopwatch & Return time recorded in milli-seconds \\
		\hline
		\textbf{13} & ThreadQueue & Get the collection of queued threads \\
		\hline
		\textbf{14} & WordCount & Split a texttt{String} by delimiter \\
		\hline
		\textbf{15} & XML Utils & Serialize a XML node into a string \\
		\hline
	\end{tabular}
	\label{tab:table3}
\end{table}


\vspace{5 pt} \noindent  \textbf{Measuring search accuracy.} We also consider multiple ways in which these various definitions of equivalence can be used to measure search accuracy; some of our ideas follow related work \cite{deep_code_search, CodeHow}. Let us assume that we have a total of $Q$ independent queries and $\textbf{res}$ be a vector of size $Q$ containing the identities of the method bodies we are searching in the database. Let us assume $\textbf{Ans}$ be a matrix of size $Q \times K$, which denotes the identity of the results returned by our system for each of those queries within a pre-defined rank $K$. We consider three metrics, out of which two metrics are based of the notion of FRank. For a particular search query $q$, $\textrm{FRank}_q$ is the smallest rank $k$ at which the user finds a code equivalent to the desired result among the top programs. 
$$\textrm{FRank} (\textbf{res}_q, \textbf{Ans}_q)  = \textrm{argmin}_k [ I(\sProg_{\textbf{res}_q} \equiv \sProg_{\textbf{Ans}_{q,k}})] $$
\noindent Here, $I$ accepts a boolean value and returns one if it is true, zero if false. Given this, our metrics are:

\vspace{5 pt}
\noindent (1) \textit{SuccessRate@K}, which estimates the probability of finding the intended result within a pre-defined rank $K$:
$$\textrm{SuccessRate@K} = \frac{1}{Q}\sum_{q}[I(\textrm{FRank} (\textbf{res}_q, \textbf{Ans}_q) \leq K)].$$

\noindent (2) \textit{Precision@K}, which estimates the fraction of the top $K$ results that are correct. Then:
$$\textrm{Precision@K} = \frac{1}{KQ}\sum_{q}\sum_{k}[I(\sProg_{\textbf{res}_q} \equiv \sProg_{\textbf{Ans}_{q,k}})]$$
\noindent (3) \textit{MRR} or Mean Reciprocal Ratio, which is simply the average inverse FRank:
$$\textrm{MRR} = \frac{1}{Q}\sum_{q}[\frac{1}{\textrm{FRank} (\textbf{res}_q, \textbf{Ans}_q) }]$$

\noindent For each metric, a larger value means higher search accuracy.





\vspace{5pt}
\subsubsection{Results and Discussion} \label{sec:automated}
\vspace{5pt}
Results for each of the four equivalence metrics - SuccessRate@1, SucccessRate@10, Precision@10, and MRR, for each of the four search methods tested, are shown in Table \ref{comparison_matrix}. 
The results suggest that, if the correct method body is available, using \textsc{Codec} we can expect to obtain an exact match 11\% of the time. We feel that this is a quite impressive result, given that \textsc{Codec} is able to select the correct method body from among 27.9M candidate methods, using only contextual information as well as the JavaDoc comment.  The chance of obtaining a ``correct'' program increases to 23\% if success is measured in terms of returning a code that can produce the same set of API call sequences, and to 27\% if success is measured in terms of obtaining a program with the same set of API calls.

It is interesting that the competitive methods fared so poorly compared to \textsc{Codec}. No other search methodology had a non-zero exact-match success rate for the top result returned. But even beyond exact-match, \textsc{Codec} dominates all other methods, over all of the various metrics.

\begin{figure*}[t]
	\centering
	\begin{subfigure}{0.33\textwidth}
		\includegraphics[width=5cm,height=4.2cm,trim={0.3cm  0.45cm 0.3cm 0.3cm},clip]{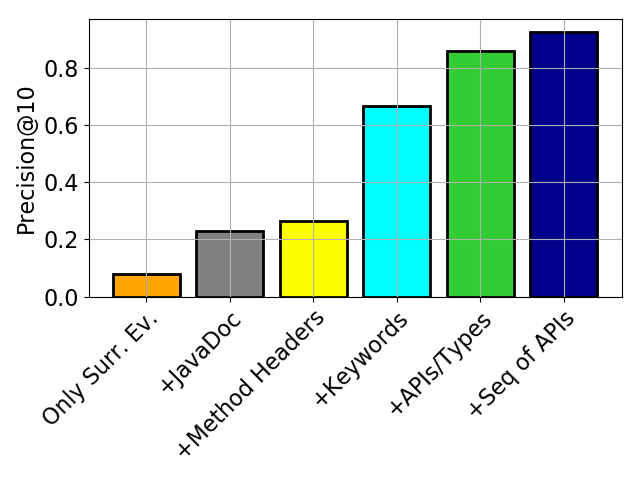}
		\caption{API Match}
	\end{subfigure} 
	\begin{subfigure}{0.33\textwidth}
		\includegraphics[width=5cm,height=4.2cm,trim={0.4cm  0.45cm 0.3cm 0.3cm},clip]{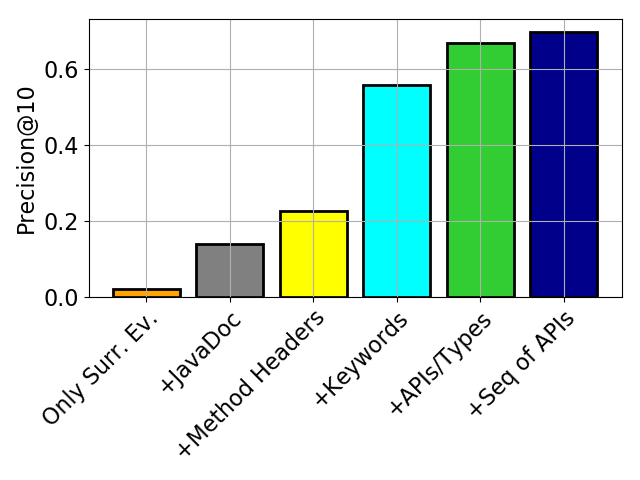}
		\caption{Sketch Match}
	\end{subfigure} 
	\begin{subfigure}{0.33\textwidth}
		\includegraphics[width=5cm,height=4.2cm,trim={0.4cm  0.45cm 0.3cm 0.3cm},clip]{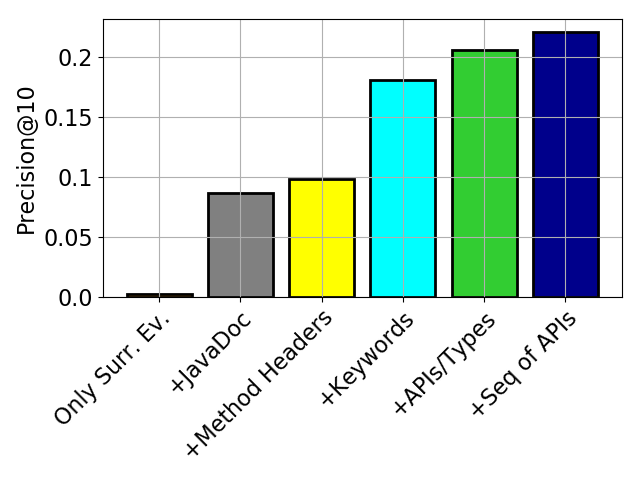}
		\caption{Exact Match}
	\end{subfigure}
	\vspace{-3 pt}
	\caption{Contributions of various types of evidence to \textsc{Codec}'s retreival accuracy.}
	\vspace{-5 pt}
	\label{fig:evidence_precision}
\end{figure*}

\vspace{5 pt}
\noindent \textbf{How useful is external context?}
In a sense, it may not be surprising that \textsc{Codec} outperforms both CodeHow and Deep-Code search, as both of these are natural-language based, and hence they focus mostly on the JavaDoc comments (though as described, we did try to augment these methods to take into account contextual information as well, at least in a cursory manner).  However, it turns out that the explanation for why 
\textsc{Codec} has much higher accuracy is not as simple as ``it uses more information to power search.''  To examine this, we repeated the experiment using \emph{only} contextual information external to the method we are searching for (that is, no JavaDoc, and no header for the method being searched for).  We then added in JavaDoc comments, the method header, and continued to add information about the internals of the method, such as the API calls present (those internals are not used in the results of Table \ref{comparison_matrix}).  Results are shown in Figure \ref{fig:evidence_precision}.  What we find is that although context and the method header do seem to add significantly to the search quality, unless one uses method internals, JavaDoc comments provide for the bulk of the accuracy. Note that JavaDoc comments are available to all competitive methods.

\begin{table}
	\centering
	\caption{$p$-value at which the hypothesis $H_0^{A,B}$ was rejected during the user study. }
	\vspace{- 3 pt}
	\small
	\begin{tabular}{|l|c|c|c|c|}
		\hline
		\textit  & CodeHow &  Deep-Code & Non-Prob & \textsc{Codec}\\
		\hline
		\hline
		CodeHow & N/A & 0.7160 & 0.0749 & \textbf{0.9584} \\
		\hline
        Deep-Code & 0.2477 & N/A  & 0.0224 & \textbf{0.9378} \\
		\hline
		Non-Prob & 0.9058 & 0.9699 & N/A & \textbf{0.9871} \\
		\hline
		\textsc{Codec} & \textbf{0.0306} & \textbf{0.0457} & \textbf{0.0084} & N/A \\
		\hline
	\end{tabular}
	\vspace{- 3 pt}
	\label{tab:table4}
\end{table}

\vspace{5 pt}
\noindent \textbf{Influence of method internals.} We see a major jump in accuracy while including evidences such keywords, API calls, and types from within the body of the method, which may be available in a partially-completed or completed method whose body is used to power search. Figure \ref{fig:evidence_precision} shows that if such evidences are available, we see a significant increase in accuracy, from about 20\% (in terms of sketch match) for only external evidences, up to nearly 70\% if a complete set of within-method evidences are supplied.

\vspace{5 pt}
\noindent \textbf{Comparison with non-probabilistic \textsc{Codec}.}
The results so far suggest that the presence of more data is not the only reason for \textsc{Codec}'s success.  To dive deeper into this, we consider in detail how \textsc{Codec} compares with its non-probabilistic version.  
Going into our experiments, we suspected that the regularization provided by the prior on $Z$ would be useful; because $\sZ$ is not known during training, any $\sZ$ value in a neighborhood must have a reasonable likelihood of decoding into the correct sketch $\sY$.  Intuitively, this will force programs that are embedded close to one another to have a reasonable similarity in terms of \textsc{Sketch} syntax, as they must share likely $\sZ$ values. This should help boost generalization ability, and hence accuracy.  In comparison, the non-probabilistic version of \textsc{Codec} simply attempts to co-locate embedded sketches and context, which may result in very weak generalization ability. However, we did not anticipate the extent to which this distinction was crucial.  

To examine this in a bit more detail, we randomly selected 10,000 methods and encoded the sketch for each of the methods into the latent space using the learned reverse encoder for \textsc{Codec}, as well as the equivalent encoder for the non-probabilistic version of \textsc{Codec}. We then clustered those 10,000 embeddings for both \textsc{Codec} and its non-probabilistic equivalent using $k$-means, with $k = 10$.  For each cluster, we measured the average Jaccard similarity of the API calls made by the methods within the cluster to the calls made by the methods within the other clusters.

\begin{figure*}[!t]
	\centering
	\begin{subfigure}[b]{0.33\textwidth}
		\includegraphics[width=5.3cm,height=4.5cm,trim={0.8cm 0.5cm 0.4cm 0cm},clip]{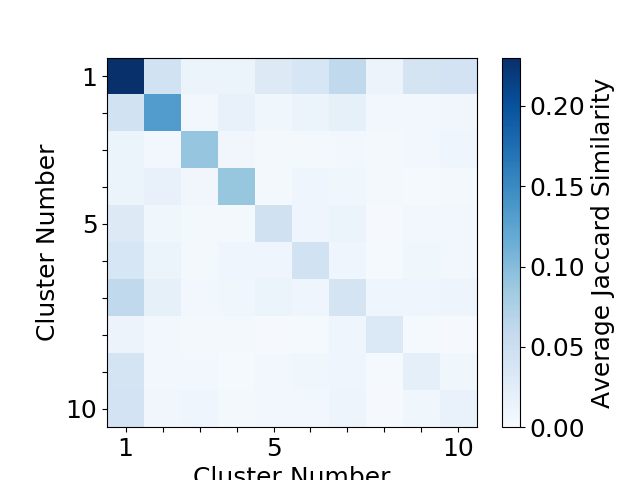}
		\caption{\textsc{Codec} inter/intra-cluster similarity.}
	\end{subfigure} 
	\begin{subfigure}[b]{0.33\textwidth}
		\includegraphics[width=5.3cm,height=4.5cm,trim={0.8cm 0.5cm 0.4cm 0cm},clip]{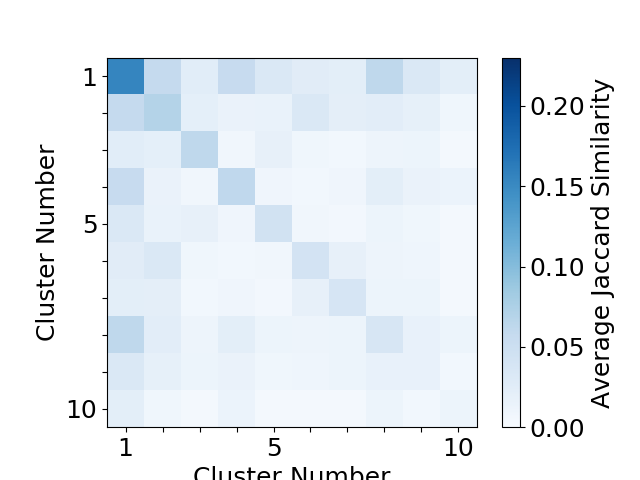}
		\caption{Non-prob inter/intra-cluster similarity.}
	\end{subfigure} 
	\begin{subfigure}[b]{0.33\textwidth}
		\includegraphics[width=5.2cm,height=4.15cm,trim={0cm 1cm 0cm 0cm}]{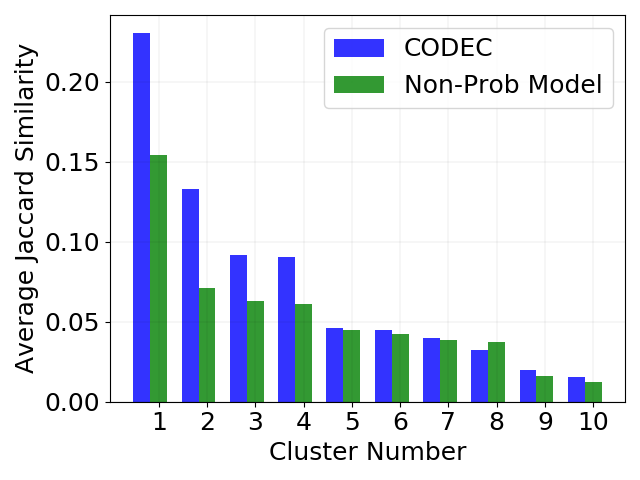}
		\caption{Comparison of inter-cluster similarity.}
	\end{subfigure}
	\caption{Comparing inter-cluster similarity (measured via API call similarity) for \textsc{Codec} and its non-probabilistic variant.}
	\vspace{-5 pt}
	\label{non_prob_compare}
\end{figure*}



As shown in Figure \ref{non_prob_compare}, what we find is that the methods within the clusters formed by \textsc{Codec} show much greater similarity compared to the clusters formed by its non-probabilistic version. The self-similarity within each cluster tops out at around 16\% for the non-probabilistic version, whereas it tops out at around 24\% for \textsc{Codec}.  This provides strong evidence that the embeddings learned by \textsc{Codec} are high-quality precisely because of \textsc{Codec}'s synthesis-based code search.

\subsection{Qualitative User Study}

Clearly, \textsc{Codec} demonstrates utility when a method body is removed, and a search methodology is able to find the method body (or an equivalent method body) in a database of millions of alternatives.  However, in reality, a correct method body is not typically available in a database, and questions of how useful a returned code is are likely best answered by human programmers.  Thus, we conducted a user study where programmers were asked to grade the utility of the programs returned by the various methods

\begin{figure*}[t]
	\centering
	 \captionsetup[subfigure]{width=0.9\textwidth}
	\begin{subfigure}[t]{0.33\textwidth}
		\includegraphics[width=5cm,height=4cm,trim={0cm 0cm 0cm 0cm},clip]{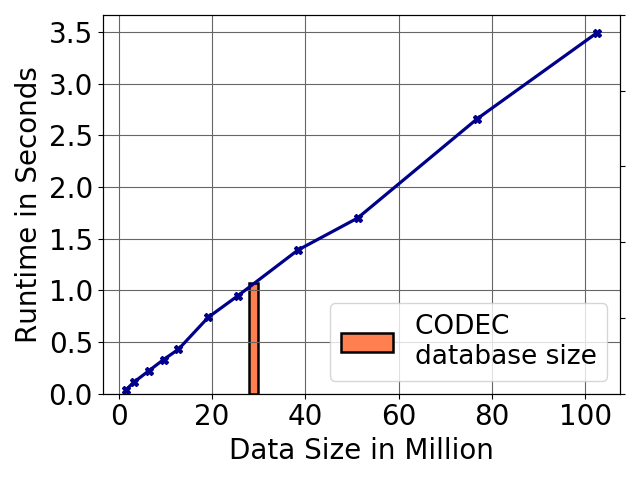}
		\caption{ Runtime variation with increasing data size for a 16-GPU machine. \textsc{Codec} uses an indexed database of 27.9M programs collected from Github. }
	\end{subfigure} 
	\begin{subfigure}[t]{0.33\textwidth}
		\includegraphics[width=5cm,height=4.1cm,trim={0cm 0cm 0cm 0cm},clip]{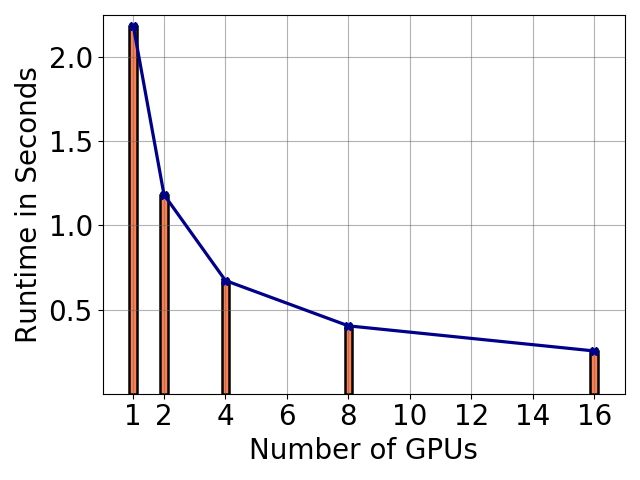}
		\caption{Runtime variation with increasing number of GPUs for a dataset of 6.4M indexed programs.} 
	\end{subfigure} 
	\begin{subfigure}[t]{0.33\textwidth}
		\includegraphics[width=5cm,height=4cm,trim={1.3cm 0.42cm 0.5cm 0.4cm}]{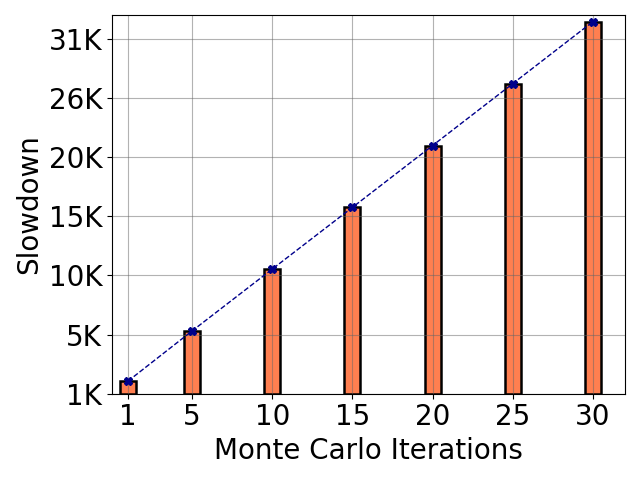}
		\caption{ Slowdown using Monte-Carlo sampling, compared to \textsc{Codec}'s reverse encoder.}
	\end{subfigure}
	\vspace{-10 pt}
	\caption{Performance characteristics of \textsc{Codec}.}
		\vspace{-5 pt}
	\label{fig:runtime_graphs}
\end{figure*}

\subsubsection{Experimental Setup}

\vspace{5 pt}
\noindent \textbf{Search problems tested}.
We constructed a set of 15, carefully created search problems, as shown in Table \ref{tab:table3}. Each search problem consists of a well-documented, hand-written class with two or more methods where one of the method bodies is missing.

\vspace{5 pt}
\noindent \textbf{Volunteers and rating instructions}.
We recruited 13 volunteers to rate search results.  Each volunteer was a Rice Computer Science graduate student, and could be described as an expert programmer.

For each search task, and for each of the four competitive methods tested, each volunteer was shown the top three search results. For each search task, users were asked to rate the set of three results as a group on a scale of one-to-five in terms of retrieved code's perceived utility for helping a programmer to fill in the missing method body (five being ``perfect match'' and one being ``poor match''). Beyond that, each volunteer was asked to develop his/her own interpretation of search result quality by examining the incomplete Java class. This amounted to 60 search result rating tasks in all, per volunteer.  The complete set of rating tasks was designed to be completed in one hour, but volunteers were not given a time limit.

\vspace{5 pt}
\noindent \textbf{Statistical analysis}.
  An average search result rating was computed for each of the four methods, across each of the 15 search problems $\times$ 13 volunteer $=$ 195 search results.  

We were also interested in the statistical significance of comparisons of the average ratings across search methodologies: If one method has a higher rating on average, is the difference statistically significant?  We designed a bootstrap-based pairwise hypothesis test \cite{bootstrap} comparing the ratings given by a user for predictions on the same problem from different algorithms.
We consider a null hypothesis of the following form.  For two search methods $A$ and $B$, define:

\vspace{8 pt}
\noindent
$H_0^{A, B} =$ ``The average score for search strategy $A$ is worse than the score for search strategy $B$.''

\vspace{8 pt}
\noindent  Our goal is to see if we can reject this null hypothesis for various combinations of $A$ and $B$.  Unfortunately, our experimental setup is rather complex, as there are two sources of variability in our experimental setup: (1) the set of participants chosen, as well as (2) the set of search tasks selected.   With a different set of participants and a different set of search tasks, we may have obtained different results.  

Hence, a classical statistical test such as a $t$-test is not directly applicable, as it assumes the ratings are sampled (identically and independently distributed) from a single population. In our case, this assumption does not hold as the scores obtained by a single volunteer are conditioned upon the volunteer selected.
Thus our use of a bootstrap-based method to attempt to reject the null hypothesis.  For a particular pair of search methods, we re-sample with replacement from among the volunteers, and we re-sample with replacement from among the rating tasks, and compute the mean score.  If $A$ has a better average than $B$, then for that re-sampled data set instance, the null hypothesis has been rejected.  This process is repeated many times, and the fraction of the time that the null hypothesis is not rejected is the $p$-value of the test.

\subsubsection{Results and Discussion}
\vspace{5pt}
Across all rating tasks and volunteers, we see that \textsc{Codec} receives an average rating of 3.89, Deep-Code search and CodeHow receive an average rating of 2.81 and 2.53 respectively, while the non-probabilistic version of \textsc{Codec} receives a rating of 1.71.\\

Qualitatively, it seems that these results show a very large spread, with \textsc{Codec} more than a full rating point higher than Deep-Code search and CodeHow, and the non-probabilistic version of \textsc{Codec} a full rating point behind Deep-Code search.  We would venture to say that for a real code-search application, these gaps would translate to significant differences in user satisfaction with the various methods, and perhaps even to gaps in programmer productivity as a user needs to spend extra time and effort with a search where she/he is not happy with the results.\\

Statistically, there are significant differences among the methods, as shown in Table \ref{tab:table4}.  \textsc{Codec}'s average score is, statistically speaking, significantly higher than the average score of the other three competitive methods---the only possible exception is \textsc{Codec} compared to Deep-Code search, where the null hypothesis is rejected at a $p$-value of $0.04$.   There is not a statistically-significant difference between Deep-Code search and CodeHow, but all methods are better than the non-probabilistic version of \textsc{Codec}.

\vspace{8pt}
\subsection{Runtime Performance}
\vspace{5pt}

One of the key technical innovations of our approach is the introduction of the reverse encoder $Q(Z|\sY, \M)$ which makes it possible to evaluate $P(\sY_i|\sX, \M)$ for a particular sketch $\sY_i$ analytically, using a closed-form formula.  This is crucial as it allows a large database to be searched quickly. In this subsection, we benchmark our \textsc{Codec} implementation, running it on an Amazon AWS \texttt{p2.16xlarge} machine with 16 NVIDIA K80 GPUs.


\textsc{Codec} uses this approximation to perform parallel/distributed search using multiple GPUs, possibly spread over multiple machines. The implementation is fairly simple. $b_{\sX}$ is pre-computed for each database program and stored in GPU RAM. In the case of a multi-dimensional latent space, each $b_{\sX}$ is a vector, whose dimensionality is equivalent to the dimensionality of the latent space (256 dimensions). Then, in response to a query, the computations of Section 7 are performed on each GPU, resulting in an approximation of $P(\sY_i | \sX)$ for each $\sY_i$ in the database.  Assuming the goal is to return the top $K$ programs, the top $K$ $P(\sY_i | \sX)$ values are sent from each GPU to a central server, where the top $K$ $P(\sY_i | \sX)$ values overall are computed, and the associated codes are returned.

\vspace{8 pt}
\noindent \textbf{Runtime with varying database size}. We start our analysis with a small database of 1.6M Java methods, where each GPU is assigned a data size of 0.1M methods, and increase the database size to a total of 102.4M synthetic programs, and measure the time to compute the top $K$ programs for $K = 100$. Note that \textsc{Codec} uses a set of 27.9 million Java methods collected from Github, which takes approximately 1.14 seconds to search. Results are shown in Figure \ref{fig:runtime_graphs}(a). Runtime increases linearly with  data size, and for a dataset of 102.4M programs, we take around 3.5 seconds per query using 16 GPUs. Considering that an Amazon \texttt{p2.16xlarge} costs \$14.40 per hour, this equates to a cost of around 1.4 cents per 100M programs searched.

\vspace{8 pt}
\noindent \textbf{Runtime with varying hardware}. Since \textsc{Codec} search is embarrassingly parallel, it should be possible to push down the runtime by simply increasing the number of GPUs available. For this experiment, we fix the dataset size to be 6.4M programs, small enough so that the associated vectors can all stored on a single GPU. We then increase the number of GPUs by repeatedly doubling from 1 until 16 GPUs, and notice that the runtime decreases approximately by half with each doubling, as shown in Figure \ref{fig:runtime_graphs}(b). This implies that it should be possible to push the search time for a large database down nearly arbitrarily, by simply using more GPUs.



\vspace{8 pt}
\noindent \textbf{Efficiency and accuracy of the reverse encoder}. An alternative to using reverse encoder (evaluated in all of the experiments this far) is to use Monte-Carlo (MC) simulation. The obvious, MC method for evaluating $P(\sY_i|\sX, \M)$ and estimating this value, is to draw $N$ samples from $P(Z|\sX, \M)$, and then use the estimator  
$P(\sY_i|\sX, \M) \approx \frac{1}{N} \sum_{\sZ_j \sim P(Z|\sX, \M)} P(\sY_i|\sZ_j, \M)$. Note that each sample requires a probability calculation, $P(\sY_i|\sZ_j, \M)$, which will be quite expensive (even on a GPU) as this probability calculation makes use of a tree-based, recurrent neural network. 


We compare \textsc{Codec}'s reverse encoder-based implementation with this MC alternative, running on a single GPU, using 0.1M programs on a \texttt{p2.xlarge} Amazon machine.

Results are plotted in Figure \ref{fig:runtime_graphs}(c), where the ratio of the MC time to \textsc{Codec}'s reverse encoder-based time is shown as a function of the number of MC iterations performed.  We find that after around 30 MC iterations, the Jaccard similarity between the top 100 programs returned by both computations for an arbitrary query converges to around 0.91, and that further MC iterations do not increase this similarity further (at this stage, the mean co-efficient of variation for the MC estimate is around 1\%). This indicates that running 30 MC iterations is a good rule-of-thumb for the MC approach, at least for our database. At this point, the MC method is around $\sim31,000$ times slower than \textsc{Codec}'s reverse encoder.

\section{Conclusions}
We have proposed the problem of contextualized code search, where a database of code fragments is searched
for a match to a query composed of various evidences extracted from the surrounding program.
The benefit of contextualized code search compared
to other code search methods is that search happens ``for free'' using the surrounding context; the user need not specify the parameters
for search.
We have proposed a general, probabilistic framework that allows the inclusion of various types of evidence
(sets of types that appear in the surrounding code, lists of formal parameters, English comments, etc.).  Virtually any evidence
can be used, as long as 
a suitable encoder for the evidence can be developed.  
A key technical innovation is the learning of a ``reverse encoder'' that allows for 
fast search, by allowing the framework to compute a simple, closed-form version of the posterior probability of generating a code from the evidence at query time.  We have shown that the resulting search engine
gives high-quality results. 

We end the paper by asking, could \textsc{Codec} be extended past Java?  
In terms of engineering effort, adding an additional language would require (a) designing a new intermediate language (similar to \textsc{Sketch}) for the target language, (b) re-implementing the context extractor and decompiler, and (c) updating the learner and search engine to incorporate any changes in evidence types and in the intermediate language. For most modern imperative languages (Python and C++ come to mind), the engineering effort would be minimal, as the evidences would stay the same, and \textsc{Sketch} could be used with small changes. However, even a functional language such as Scala should require relatively little effort.

Perhaps a more interesting question is: would \textsc{Codec} give good results with other languages?  We anticipate it would, with one caveat: our \textsc{Codec} prototype relies heavily on the fact that Java has a widely-used set of standard types and methods.  These are important evidence types for \textsc{Codec}.  In a language such as C for which there is arguably less uniformity in terms of the types and libraries used, the \textsc{Codec} approach might be more successful for searching a more limited code base (say, the code produced by a corporation or open-source project), as opposed to searching a more general database such as GitHub. 

\section{Acknowledgements}
We thank the anonymous reviewers for their constructive  suggestions.  We thank the volunteers for their time in helping us with the qualitative user study experiment. Work presented in this paper has been supported by NSF under grant numbers 1910803 and 1918651.

\section{Appendix: Reverse Encoder in Multi Dims}

The goal is to be able to compute $P(\sY|\sX)$ in the case that $\sZ$, the embedding, is multi-dimensional. We begin with:
\begin{equation*}
\begin{aligned}
P(\sY|\sX) &= \int_\sZ P(\sZ|\sX)P(\sY|\sZ) d\sZ \\
& = \int_\sZ \frac{P( \sZ|\sX)P(\sZ|\sY)P(Y)}{P(\sZ)} d\sZ \\
& = \int_\sZ \exp \big( \textbf{a}_{\sX} \cdot \sZ^2 + \textbf{b}_{\sX} \cdot \sZ + \textbf{c}_{\sX} \big) \\ 
& \ \ \times \exp \big( \textbf{a}_{\sY} \cdot \sZ^2 + \textbf{b}_{\sY} \cdot \sZ + \textbf{c}_{\sY} \big) \\
& \ \ \times \exp \big( -\textbf{a}_{\sI} \cdot \sZ^2 - \textbf{b}_{\sI} \cdot \sZ- \textbf{c}_{\sI} \big) \times P(Y) d\sZ\\
\end{aligned}
\end{equation*}

\noindent where $(\textbf{a}_{\sX}, \textbf{b}_{\sX}, \textbf{c}_{\sX} )$, $(\textbf{a}_{\sY}, \textbf{b}_{\sY}, \textbf{c}_{\sY} )$, $(\textbf{a}_{\sI}, \textbf{b}_{\sI}, \textbf{c}_{\sI} )$ are the parametarizations of multidimensional normal distributions $P(\sZ|\sX)$, $P(\sZ|\sY)$ and $P(\sZ)$, respectively.

Note that each of these can be represented as follows, for each dimension $i$:
\begin{equation*}
\begin{aligned}
&\textbf{a}_{\sX}^i = -\frac{1}{2\sigma_{\sX}^2}, \textbf{b}_{\sX}^i = \frac{\mu^i_\sX}{\sigma_{\sX}^2}, \textbf{c}_{\sX}^i = -\frac{(\mu_\sX^i)^2}{2\sigma_{\sX}^2} - \frac{1}{2}\ln(\sigma_{\sX}^2) - \frac{1}{2}\ln2\pi \\
&\textbf{a}_{\sY}^i = -\frac{1}{2\sigma_{\sY}^2}, \textbf{b}_{\sY}^i = \frac{\mu^i_\sY}{\sigma_{\sY}^2}, \textbf{c}_{\sY}^i = -\frac{(\mu_\sY^i)^2}{2\sigma_{\sY}^2} - \frac{1}{2}\ln(\sigma_{\sY}^2) - \frac{1}{2}\ln2\pi \\
&\textbf{a}_{\sI}^i = -\frac{1}{2}, \ \ \ \ \ \textbf{b}_{\sI}^i = 0, \ \ \ \ \textbf{c}_{\sI}^i = - \frac{1}{2}\ln2\pi \\
\end{aligned}
\end{equation*}
\noindent Continuing, we have:
\begin{equation*}
\begin{aligned}
P(\sY|\sX) &= P(\sY) \times \int_\sZ \exp \big[ (\textbf{a}_{\sX} + \textbf{a}_{\sY} - \textbf{a}_{\sI} ) \cdot \sZ^2 + \\
&\ \ \ \ \ \ \ + (\textbf{b}_{\sX} + \textbf{b}_{\sY}) \cdot \sZ + ( \textbf{c}_{\sX} + \textbf{c}_{\sY} - \textbf{c}_{\sI} ) \big] d\sZ\\
&= P(\sY) \times \int_\sZ \exp \big( \textbf{a}^* \cdot \sZ^2 + \textbf{b}^* \cdot \sZ + \textbf{c}^\prime \big) d\sZ\\
\end{aligned}
\end{equation*} 

\vspace{2 pt}

\noindent Where:
$\textbf{a}^*= \textbf{a}_{\sX} + \textbf{a}_{\sY} - \textbf{a}_{\sI}$,\ \  \  $\textbf{b}^*= \textbf{b}_{\sX} + \textbf{b}_{\sY}$ and 
$\textbf{c}^\prime= \textbf{c}_{\sX}+ \textbf{c}_{\sY} - \textbf{c}_{\sI}$. 

\noindent Simplifying, we have:
\begin{equation*}
\begin{aligned}
P(\sY|\sX) &= P(\sY) \times \exp( \textbf{c}^\prime - \textbf{c}^*) \\
& \ \ \ \ \times \int_\sZ \exp(\textbf{a}^* \cdot \sZ^2 + \textbf{b}^* \cdot \sZ + \textbf{c}^*) d\sZ\\
& = P(\sY) \times \exp(\textbf{c}^\prime-\textbf{c}^*) \\
\end{aligned}
\end{equation*}

\vspace{-10 pt}

\noindent where,
\begin{equation*}
\begin{aligned}
\textbf{c}^* &= \frac{\textbf{b}^{*2}}{4\textbf{a}^*} + \frac{1}{2}\ln(-\frac{\textbf{a}^*}{\pi}) \\
\textbf{c}^\prime \ &= \textbf{c}_{\sX} + \textbf{c}_{\sY} - \textbf{c}_{\sI}  \\
& = \frac{\textbf{b}_{\sX}^2}{4\textbf{a}_\sX} + \frac{1}{2}\ln(-\frac{\textbf{a}_\sX}{\pi}) +  \frac{\textbf{b}_\sY^2}{4\textbf{a}_\sY} + \frac{1}{2}\ln(-\frac{\textbf{a}_\sY}{\pi}) -\frac{1}{2} \ln2\pi \\
&= \frac{\textbf{b}_{\sX}^2}{4\textbf{a}_\sX} + \frac{\textbf{b}_\sY^2}{4\textbf{a}_\sY} + \frac{1}{2}\ln(-\frac{\textbf{a}_\sX}{\pi}) + \frac{1}{2}\ln(-\frac{\textbf{a}_\sY}{\pi}) -\frac{1}{2} \ln2\pi \\
\end{aligned}
\end{equation*}

\noindent Finally this reduces our final computation to,
\begin{equation*}
\begin{aligned}
& \log P(\sY|\sX) = \log P(\sY) + \sum_i(\textbf{c}^\prime_i - \textbf{c}^*_i) \\
\end{aligned}
\end{equation*}

\vspace{-10 pt}
\noindent where,
\begin{equation*}
\begin{aligned}
\textbf{c}^\prime - \textbf{c}^* = &\frac{\textbf{b}_{\sX}^2}{4\textbf{a}_\sX} + \frac{\textbf{b}_\sY^2}{4\textbf{a}_\sY} - \frac{\textbf{b}^{*^2}}{4\textbf{a}^*} + \frac{1}{2}\ln(-\frac{\textbf{a}_\sX}{\pi}) + \\
&\frac{1}{2}\ln(-\frac{\textbf{a}_\sY}{\pi}) +
- \frac{1}{2}\ln(-\frac{\textbf{a}^*}{\pi}) -\frac{1}{2} \ln2\pi
\end{aligned}
\end{equation*}

Note that this computation is easily implemented to a matrix-vector style operation that can be parallelized to run on a GPU.

\balance

\newpage

\bibliographystyle{abbrv}
\bibliography{code_search}

\newpage
\balance

\end{document}